\documentclass[10pt,preprint2]{aastex}
\usepackage{graphicx}
\usepackage{epstopdf} 

\begin{document}

\title{The Circumstellar Disk of the Be Star $o$~Aquarii}

\author{
T.\ A.\ A.\ Sigut,\altaffilmark{1}$^{,}$\altaffilmark{2}
C. Tycner,\altaffilmark{3}
B.~Jansen,\altaffilmark{3}
R.~T.~Zavala\,\altaffilmark{4}
\\
\medskip
{\small Accepted for publication in the Astrophysical Journal}
}

\altaffiltext{1}{Department of Physics and Astronomy, The University of
Western Ontario, London, Ontario, Canada N6A 3K7}
\altaffiltext{2}{Centre for Planetary Science and Exploration, The University of
Western Ontario, London, Ontario, Canada N6A 3K7}
\altaffiltext{3}{Department of Physics, Central Michigan University,
  Mount Pleasant, MI 48859, USA}
\altaffiltext{4}{US Naval Observatory, Flagstaff Station, 10391
  W.~Naval Observatory Rd., Flagstaff, AZ 86001}

\begin{abstract}

Omicron Aquarii is late-type, Be shell star with a stable and
nearly symmetric H$\alpha$ emission line. We combine H$\alpha$
interferometric observations obtained with the Navy Precision Optical
Interferometer (NPOI) covering 2007 through 2014 with H$\alpha$
spectroscopic observations over the same period and a 2008 observation
of the system's near-infrared spectral energy distribution to
constrain the properties of $o$~Aqr's circumstellar disk.  All
observations are consistent with a circumstellar disk seen at
an inclination of $75\pm\,3^{\circ}$ with a position angle on the
sky of $110\pm\,8^{\circ}$ measured E from N. From the best-fit disk
density model, we find that 90\% of the H$\alpha$ emission arises
from within $9.5$ stellar radii, and the mass associated with this
H$\alpha$ disk is $\sim 1.8\times10^{-10}$ of the stellar mass and the
associated angular momentum, assuming Keplerian rotation for the disk,
is $\sim 1.6\times10^{-8}$ of the total stellar angular momentum. The
occurrence of a central quasi-emission (CQE) feature in Mg\,{\sc ii} $\lambda\,4481$
is also predicted by this best-fit disk model and the computed profile compares successfully 
with observations from 1999.
To obtain consistency between the H$\alpha$
line profile modelling and the other constraints, it was necessary
in the profile fitting to weight the line core (emission peaks and central depression) more
heavily than the line wings, which were not well reproduced by 
our models. This may reflect the limitation of assuming a single
power-law for the disk's equatorial density variation.
The best-fit disk density model for $o$~Aqr predicts that H$\alpha$ is near
its maximum strength as a function of disk density, and hence the H$\alpha$
equivalent width and line profile change only weakly in response to
large (factor of $\sim 5$) changes in the disk density. This may in part 
explain the remarkable observed stability of o~Aqr's H$\alpha$ emission line profile.

\end{abstract}

\keywords{techniques: interferometric -- stars: circumstellar matter
  -- stars: emission line, Be -- stars: individual ($o$~Aqr)}

%
%

\section{Introduction}
\label{sec:intro}

$o$~Aqr (HR 8402, HD 209409) is a bright, Be shell star of spectral
type B7IVe.  \cite{riv06} note that $o$~Aqr has had stable H$\alpha$
emission and does not exhibit $V/R$ variations, thus excluding a
prominent one-armed spiral density wave in the disk
\citep{oka91,han95}.  However, $o$~Aqr does possess a central
quasi-emission feature (CQE) in Mg\,{\sc ii} $\lambda\,4481$ \cite{riv06},
consistent with a high viewing angle for the disk
\citep{han96}. \citet{sto87} used the shape and widths of He\,{\sc i}
$\lambda\,4471$ and Mg\,{\sc ii} $\lambda\,4481$ and a gravitational
darkening model for the central star to estimate the inclination
angle for $o$~Aqr, finding $i>82^{\circ}$, consistent with its shell
designation. \citet{hub09} claimed detection of a weak magnetic field
in $o$~Aqr of about 100~G at 3$\sigma$; however, \citet{bag12}
conclude that the polarization detected with the FORS1~VLT instrument
was instrumental in nature. $o$~Aqr is not known to have a binary
companion, a result strengthened by direct AO imaging observations in
the K-band with VLT \citep{oud10}. The $v\sin i$ of $o$~Aqr is
$282\,\rm km\,s^{-1}$, giving a $V/V_{\rm crit}$ ratio of $0.74$
\citep{tou13}. This is consistent with the consensus that rapid
rotation is a key driver behind the Be phenomena \citep{how07,riv13b,
  riv13a}.

As both bright ($m_{\rm V}=4.69$) and close ($d=134\;$pc; based on
{\it Hipparcos} parallax), $o$~Aqr has been a target of recent
interferometric studies. \cite{mei12} included it in their survey of Be
stars and \cite{tou13} resolved it in the K-band, finding a major axis
of $1.525\pm 0.642\,$mas as measured by fitting a geometric star-plus-Gaussian
disk to the observed visibilities.

Optical interferometry using H$\alpha$ emission has proven to be very
effective in resolving the circumstellar emission of Be stars. The
strength of H$\alpha$ can result in detectable emission extending to
many stellar radii \citep{tyc05,gru06}. The combination of H$\alpha$
interferometry and contemporaneous H$\alpha$ spectroscopy has been
shown to be a powerful tool to constrain the physical properties of Be
star circumstellar disks \citep{tyc08,jon08}. In this current work, we
attempt to constrain the physical parameters of the H$\alpha$ emitting
circumstellar disk surrounding $o$~Aqr using this approach. We
attempt to find a unified disk density model that reproduces the
observed H$\alpha$ emission profile, H$\alpha$ interferometric
visibilities, the near-IR spectral energy distribution, and the
existence of a CQE in Mg\,{\sc ii} $\lambda\,4481$, all using the {\sc
bedisk\/} \citep{sig07} and {\sc beray\/} \citep{sig11} numerical
codes.

%
%

\section{Observations}
\label{sec:obs}

\subsection{Spectroscopy}

Spectroscopic observations in the H$\alpha$ region have been obtained
using the Solar Stellar Spectrograph on the John S.\ Hall telescope at
Lowell Observatory. Thirty individual spectra are available for
observing seasons from 2005 through 2014.  The raw echelle spectral
frames have been processed using the standard reduction routines
developed by \citet{hal94} for the instrument. The H$\alpha$ profile
of $o$~Aqr is doubly-peaked, symmetric, and very stable over the time
period considered.  The stability of the H$\alpha$ emission is
important as our interferometric observations cover very similar time
period (2007 through 2014), and we combine all available visibilities into a
single analysis.

\begin{figure}
\plotone{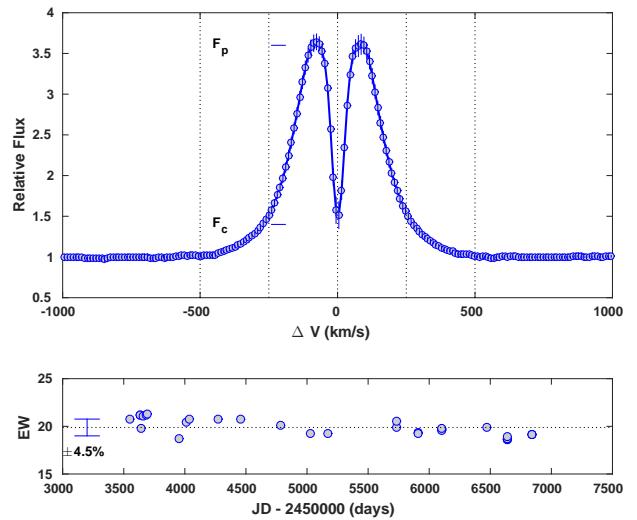}
\caption{\small Top panel: The mean H$\alpha$ line profile of $o$~Aqr
  over 2005 through 2014. The $1\sigma$ variation is shown as the error bars.
  The shell parameter, defined as $F_p/F_c$, is 2.2, and the spectral
  resolving power is $10^4$.  Bottom panel: The H$\alpha$ equivalent width
  as a function of the Julian date of the
  observations. The mean equivalent width ($19.9\,$\AA) is shown as
  the dotted line, and the $1\sigma$ variation is shown as the error
  bar.  \label{fig:halpha}}
\end{figure}

Figure~\ref{fig:halpha} shows the mean profile with the $1\sigma$
variation shown as the error bars. The lower panel of this figure also
shows the H$\alpha$ equivalent width~(EW) as a function of Julian date
over the nine year period covered by the observations. The mean EW is
$19.9\,$\AA, with a $1\sigma$ variation of only $0.9\,$\AA\ or 4.5\%.\footnote{
We have additional H$\alpha$ spectra from June 24, 2015 (JD=2457198)
which is consistent with the profile and equivalent widths of Figure~\ref{fig:halpha}.}
The profile has a peak-to-continuum contrast of 3.75, and the
H$\alpha$ shell parameter, defined as the ratio of the average flux in
the emission peaks divided by the flux at line centre, is 2.2.  This
identifies $o$~Aqr as a shell star following \citet{han96}: shell
stars have ratios in excess of $1.5$ based on the correlation of the
H$\alpha$ shell parameter with net, line-centre absorption in weak,
optically thin Fe\,{\sc ii} lines \citep{han96}. Typically for shell
stars, the viewing inclination of the system is in excess of
$70^{\circ}$.

Finally, Figure~\ref{fig:halpha} shows that the emission in 
H$\alpha$ extends to $\approx\,\pm\,400\;\rm km\,s^{-1}$. Using the
adopted mass and radius for o~Aqr from Table~\ref{tab:bparam}, the velocity at the inner edge
of a Keplerian disk is $\approx\,500\;\rm km\,s^{-1}$. As the
inclination of the system must be large and $\sin\i\approx\,1$, there is no
evidence of disk emission beyond the velocities available in the disk.
Note that this remains the case even if o~Aqr were critically rotating
(see discussion below); the disk would then start at $1.5\,R_*$
due to the geometric distortion  caused by
rapid rotation, and the velocity of the inner edge of the disk would
drop to $\approx\,400\;\rm km\,s^{-1}$.

\subsection{Interferometry}

We have acquired interferometric observations of $o$~Aqr using the
Navy Precision Optical Interferometer (NPOI) on a total of 58~nights
covering five observing seasons: 2007~Jun, 2011~Oct, 2012~Oct through Nov,
2013~Oct through Dec, and 2014~Jul.  The NPOI is a long-baseline
interferometer that can measure the fringe contrast between various
telescope pairs (i.e., baselines) for up to 6 telescopes
simultaneously~\citep{arm98}.  The fringe contrast represents the
measure of the degree of coherence between the light beams from
separate telescopes and when expressed as a squared visibility~($V^2$)
represents the normalized Fourier power of the brightness distribution
of the source on the sky~\citep{hum08}.  Therefore, assuming the
source is spatially resolved, it allows the angular extent of the
source to be constrained.

\begin{table}[t]
\begin{center}
\footnotesize
\parbox{2.9in}{
\caption{Adopted stellar parameters for $o$~Aqr \label{tab:bparam}}
}
\begin{tabular}{@{}ll}
\hline \hline
Parameter      \hspace{4cm}    &   Value \\
\hline
Mass$^{\rm a}$ ($M_{\odot}$)   & 4.2       \\
Radius$^{\rm a}$ ($R_{\odot}$)      & 3.2             \\
Luminosity ($L_{\odot}$)  & $3.6\times 10^2$ \\
$\rm T_{eff}$(K)         & $14,\!000$        \\
$\log(g)$                & $4.0$            \\
Distance$^{\rm b}$ (pc)            & 134              \\
Angular Diameter (mas)   & 0.222            \\
\hline
\end{tabular}
\medskip\\
\parbox{2.9in}{
\footnotesize
{\bf Notes.}\\
$^{\rm a}$ Adopted from \citet{tow04}.\\
$^{\rm b}$ Based on {\it Hipparcos} parallax \citep{per97}.
}
\end{center}
\end{table}

The processing of NPOI data has been conducted using the {\sc oyster}
(Optical Interferometer Script Data Reduction) package developed by
Christian Hummel, which follows the procedures outlined in
\citet{hum98} with additional bias corrections using off-fringe
measurements~\citep{hum03}. Typically, for an unresolved point source
on the sky, no loss of fringe contrast would be expected; however,
atmospheric and instrumental effects will contribute towards loss of
coherence between light beams from separate telescopes.  These effects
are typically removed from the data by interleaving the observations
of the target star with observations of a source of a known angular
diameter (i.e., a calibrator star) that allows the determination of
instrumental and atmospheric response functions, which in turn can be
divided out of the data of the target star.  However, because the
light at the beam combiner of NPOI is dispersed over 16 spectral
channels covering the wavelength range 560-870~nm, and the
H$\alpha$ emission line is contained in a single 15-nm wide spectral
channel, it is possible to calibrate the H$\alpha$ visibilities with
respect to continuum channels.  This was accomplished by adopting an
angular diameter for the central star of 0.222~mas (based on the
distance and radius listed in Table~\ref{tab:bparam}) and following
the method outlined in \citet{tyc03} with the additional step of small
channel-to-channel fixed-pattern removal~\citep{tyc06spie} that
utilized observations of two calibrator stars, $\eta$~Aqr (for 2007,
2012, and 2014) and $\iota$~Aqr (for 2011 and 2013).

\begin{figure}
\plotone{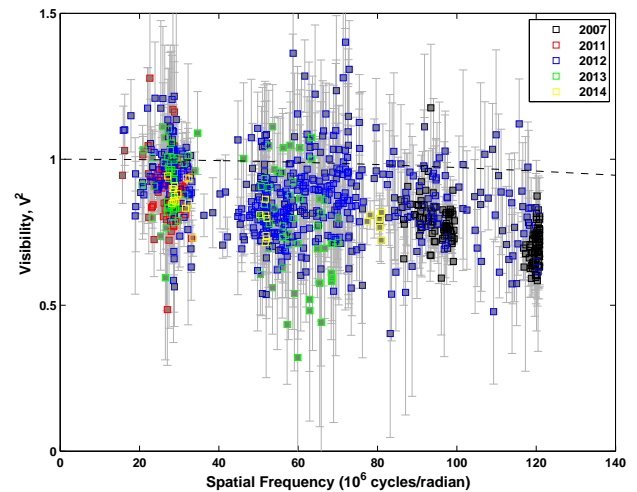}
\caption{\small NPOI squared visibilities from the H$\alpha$ channel for
  $o$~Aqr ($N=994$) as a function of the magnitude of the spatial frequency. The
  symbol colours indicate the observing seasons: black (2007), red
  (2011), blue (2012), green (2013), and yellow (2014). The signature of a central
  star as represented by a uniform disk with a diameter of
  0.222~mas is shown as the dashed-line.
\label{fig:v2}}
\end{figure}

\begin{figure}
\plotone{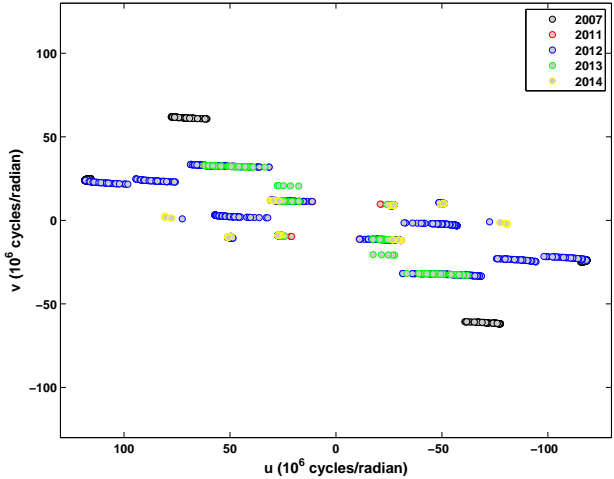}
\caption{\small The $(u,v)$ plane coverage for all NPOI
  interferometric observations.  Different colours represent individual
  observing seasons: 2007 (black), 2011 (red), 2012 (blue), 2013
  (green), and 2014 (yellow).
\label{fig:uv}}
\end{figure}

The final calibrated interferometric data set for $o$~Aqr from the
spectral channel containing the H$\alpha$ emission line consists of a
total of 994 distinct $V^2$ measurements and these are shown in
Figure~\ref{fig:v2} with the corresponding values listed in
Table~\ref{tab:visData}.  The $(u,v)$-plane coverage for the entire
data set is shown in Figure~\ref{fig:uv}. As there is good coverage in
both the N-S~($v$ spatial frequencies) and E-W~($u$ spatial
frequencies) directions, the position angle of the disk should be
reliably determined, especially as a high-axial ratio is expected from
$o$~Aqr's shell classification.

\begin{table*}[t]
\begin{center}
\small
\parbox{5in}{
\caption{The NPOI H$\alpha$ Interferometric Visibilities for $o$~Aqr$^1$ \label{tab:visData}}
}
\begin{tabular}{@{}lcccc}
\hline \hline
Julian Date       &    Spatial Frequency $u$     & Spatial Frequency $v$        &         &         \\
(JD$-2,450,000)$  &  ($10^6$ cycles rad$^{-1}$)  & ($10^6$ cycles rad$^{-1}$)   & $V^2\,\pm 1\sigma$   & Baseline$^2$ \\ \hline
4264.946  &  $  77.594$  &  $  61.891$  &  $0.757 \pm$ 0.047 & AN-W7 \\
4264.946  &  $ 115.830$  &  $  24.900$  &  $0.699 \pm$ 0.087 & E6-W7 \\
4264.990  &  $  69.858$  &  $  61.141$  &  $0.788 \pm$ 0.049 & AN-W7 \\
4264.990  &  $ 118.350$  &  $  23.709$  &  $0.749 \pm$ 0.056 & E6-W7 \\
4265.952  &  $  76.495$  &  $  61.799$  &  $0.829 \pm$ 0.112 & AN-W7 \\
\ldots    &              &              &                    &       \\
\hline
\end{tabular}
\medskip\\
\parbox{4.9in}{\footnotesize
$^1$ This table is available in its entirety only in the on-line journal. A portion is shown here
for guidance regarding its form and content.
\medskip\\
$^2$ The baseline entries for the NPOI instrument are explained in \citet{arm98}.}
\end{center}
\end{table*}

%
%

\section{Modelling}
\label{sec:model}

The parameters adopted for the central B star of the $o$~Aqr system
are given in Table~\ref{tab:bparam}. The spectral type of $o$~Aqr is
somewhat uncertain, usually quoted as either B7IVe \citep{les68} or
B6IIIe \citep{riv06}. $o$~Aqr appears in the work of \citet{fre05} who
determine the fundamental parameters for many Be stars accounting for
gravitational darkening. \cite{fre05} list ``apparent" parameters of
$T_{\rm eff} = 12,\!900 \pm 400 K$ and $\log g=3.70 \pm 0.07$ and
``pnrc" parameters (parameters of a non-rotating model which when spun
up match the star) of $T_{\rm eff} = 14,\!560 \pm 500 K$ and $\log
g=3.99 \pm 0.08$. One limitation of the \cite{fre05} work is that the
gravitational darkening formulation of \cite{col66} was used, which
seems to overestimate the gravitational darkening effect.
A recent reformulation of photospheric gravitational
darkening by \cite{elr11} gives a weaker temperature contrast between
the pole and equator and better fits the available interferometric
observations of rapidly rotating stars.  Using the \cite{elr11}
formalism and a rotation rate as a fraction of the critical rate of
$v_{\rm frac}=0.75$ \citep{mei12,tou13}, we find a pole to equator
variation of $T_{\rm eff}$ of $14,\!600\;$ to $12,\!200\;$K, versus
$15,\!000$ to $11,\!800\;$K for the \citet{col66} formalism. In
addition, the (common) Roche geometry predicts an enhancement of the
equatorial radius by $R_e/R_p=3.94/3.23=1.23$. Given that angular
diameter of the central star ($\sim 0.2\;$mas) is not expected to be
spatially resolved at the baselines utilized for the current
study~(recall Fig.~\ref{fig:v2}), we have chosen to neglect these
modest effects of gravitational darkening and have treated the central
star of $o$~Aqr as a spherical object of uniform $T_{\rm eff}$. As to
the atmospheric parameters, we have adopted the default B7 parameters
given by \citet{tow04}, which are listed in Table~\ref{tab:bparam}.

The {\sc bedisk} code of \cite{sig07} was used to compute the thermal structure
and atomic level populations of a series of equatorial, circumstellar disks with
the disk density parametrized by $(\rho_0,n)$ in the equation
\begin{equation}
\rho(R,Z) = \rho_o \left(\frac{R_*}{R}\right)^{n} \,
e^{-\left(\frac{Z}{H}\right)^2} \;.
\label{eq:rho} 
\end{equation}
Here $R$ is the radial distance from the star's rotation axis and $Z$ is the
height above or below the plane of the disk. The vertical scale height of the
disk is assumed to follow from hydrostatic equilibrium, parametrized by a single
temperature $T_0$ as
\begin{equation}
H=\left(\frac{2R_*^3\;kT_{0}}{GM_*\;\mu_m m_{\rm H}}\right)^{1/2} \,
\left(\frac{R}{R_*}\right)^{3/2}\equiv H_0\,\left(\frac{R}{R_*}\right)^{3/2}\;,
\label{eq:scale_height}
\end{equation}
where $M_*$ and $R_*$ are the mass and radius of the central star, and
$\mu_m$ is the mean-molecular weight of the gas in the disk. The
hydrostatic temperature for $o$~Aqr's disk, used only to fix the density
scale height, was set to $T_0=0.6\,T_{\rm eff} = 8,\!400\;$K \citep[see][]{sig09}. 
This choice gives $H_0/R_*=0.029$ at the inner edge of the disk.

{\sc bedisk} models were computed for ten $\rho_0$ values in
the range $10^{-12}$ to $2.5\times 10^{-10}\;\rm g\,cm^{-3}$
($\Delta\log\rho_0=0.266$) and five $n$ values from $2.0$ to $4.0$
($\Delta\,n=0.5$). These $(\rho_0,n)$ values cover the range usually found
for Be star disks based on H$\alpha$ spectroscopy \citep{sil10,sil14}.
The temperature and density structure of the disk, as well as all of
the atomic level populations, were used as input to the {\sc beray}
code \citep{sig11} which computes H$\alpha$ line profiles, SEDs, and
monochromatic images on the sky, given a specified viewing angle $i$.
{\sc beray} works by solving the equation of radiative transfer along a
series of rays directed at the observer. Rays terminating on the stellar
surface use a photospheric boundary condition, either an appropriate
stellar SED or photospheric H$\alpha$ profile Doppler shifted by the
star's projected rotation. The circumstellar disk was assumed to be
rotationally supported (i.e.\ in Keplerian rotation).

%
%

\section{Results}
\label{sec:results}

\subsection{H$\alpha$ Equivalent Width and Profile}
\label{subsec:halpha}

\begin{figure}
\plotone{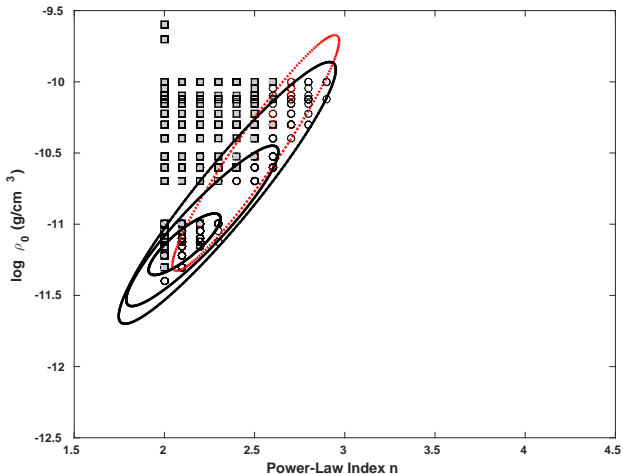}
\caption{\small Distribution of models in the $(n,\log\rho_o)$ plane that
match the observed mean H$\alpha$ EW within $2\sigma$ (grey squares). Note
that a square is plotted if any of the $R_d$ or $i$ models for a given
$n$ and $\rho_0$ satisfy the condition. The black ellipses enclose the
models (shown as open black circles) that fit the observed H$\alpha$
profile with a figure-of-merit (Eq.~\protect\ref{eq:fom}) within 10\%,
20\% and 30\% (in order of increasing ellipse size) of the best-fit model.
The dotted red ellipse encloses the models (shown as red open circles) that fit 
the observed H$\alpha$ profile with a core-weighted figure-of-merit 
(Eq.~\protect\ref{eq:fom2}) within 10\% of the best-fit model.
\label{fig:modelEW}}
\end{figure}

Several thousand H$\alpha$ line profiles were computed covering a
range in $\rho_0$, $n$, $i$, and various disk truncation
radii~($R_d$). H$\alpha$ profiles for the 50 combinations of
$(\rho_0,n)$ used in the {\sc bedisk} calculations were interpolated
down to grid spacings of $\Delta\log\rho_0=0.1$ and $\Delta\,n=0.1$.
Five disk radii were considered ($R_d=5$, $12$, $25$ and $50\,R_*$),
along with 11 viewing inclinations, covering 0 to 90~degrees, for each
model.  Figure~\ref{fig:modelEW} shows the distribution of disk
density parameters $(n,\log\rho_o)$ that match the observed mean
H$\alpha$ EW within $2\sigma$ ($19.9\,\pm\,1.8$\AA).  Note that there
are many values of $R_d$ and $i$ corresponding to each
$(n,\log\rho_o)$ pair; Figure~\ref{fig:modelEW} indicates a match if
one or more combinations of $R_d$ and $i$ match the observed
equivalent width range.

To further refine the model, a match to the H$\alpha$ line profile was
sought. A figure of merit, ${\cal F}$, for each model profile was
found by taking the average absolute fractional deviation between each
model profile (convolved to a resolving power of $10^4$) and an
observed profile:
\begin{equation}
{\cal F} \equiv \frac{1}{N}\sum_i \frac{|F_i^{\rm mod} 
                - F_i^{\rm obs}|}{F_i^{\rm obs}} \;.
\label{eq:fom}
\end{equation}
Here, $F_i^{\rm mod}$ is the model flux computed with {\sc beray},
$F_i^{\rm obs}$ is the observed flux, and
the sum over $i$ is for all $N$ wavelengths in the range
$6550\le\lambda_i\le 6570$.  The minimum in ${\cal F}$ then defined the
best-fit model.\footnote{We note that other choices for the fit figure-of-merit,
such as $(F_i^{\rm mod}-F_i^{\rm obs})^2/\sigma$, where $\sigma$ is a
wavelength-independent uncertainty, or $(F_i^{\rm mod}-F_i^{\rm obs})^2/F_i^{\rm obs}$,
select the same best-fitting models, with small permutations of the order. We
prefer the form given in Equation~\ref{eq:fom} because equal terms indicate the
same percentage deviation at each wavelength.}

\begin{figure}
\plotone{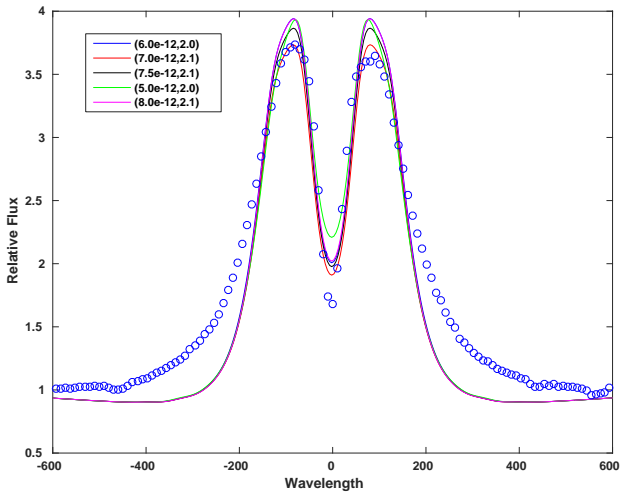}
\caption{\small The five best-fitting H$\alpha$ line profiles according
  to the figure-of-merit given by Eq.~\ref{eq:fom} (coloured lines)
  to the observed June 26, 2005 H$\alpha$ line profile (blue circles). 
  The disk parameters for each of the models (base disk density and power-law index)
  are as indicated in the legend. All models had $i=75^{\circ}$ and $R_d=25\,R_*$.
  \label{fig:bestfitHa}}
\end{figure}

Figure~\ref{fig:bestfitHa} shows the five best-fitting profiles to the
first available H$\alpha$ spectrum in our series (June 26, 2005).  A
good match to the the peak height and central depth is obtained for
the model with $\rho_0=6.0\times 10^{-12}\,\rm g\,cm^{-3}$, $n=2.0$,
$R_d=25\,R_*$ and $i=75^{o}$.  However, the computed models are all
slightly narrower at the base of the line, and the EW of the
best-fitting H$\alpha$ line profile is $16\,$\AA, somewhat less than
that of the observed profile.  Figure~\ref{fig:modelEW} also shows the
top 17 fitting profiles in the $(n,\log\rho_o)$ plane which have
${\cal F}$ within 10\% of the best-fitting model. A much narrower
region is now permitted, $-11.5 < \log\rho_0 < -11$ and $2 < n < 2.3$.
Among the top 10\% of line profile fitting models, the mode of $i$ is
$75^0$ and the mode of $R_d$ is $25\,R_*$.  The result of
$i=75^{\circ}$ is consistent with the classification of $o$~Aqr as a
shell star based on the H$\alpha$ shell parameter defined by
\citet{han96}.

Fitting all available H$\alpha$ line profiles (2005-2014) results in
disk parameters that vary only slightly relative to those found for
the June 26, 2005 profile. All best-fit parameters are identical with
the exception of $\rho_0$ which assumes values of $5.0\times
10^{-12}\,\rm g\,cm^{-3}$, (12 spectra), $6.0\times 10^{-12}\,\rm
g\,cm^{-3}$ (10 spectra), and $7.0\times 10^{-12}\,\rm g\,cm^{-3}$ (2
spectra). The average $\rho_0$, weighted by the number of
spectra,\footnote{A total of 24, and not 30, spectra are used because
  spectra taken on the same night were combined in the analysis.} is
$(5.6\pm\,0.7)\,\times 10^{-12}\,\rm g\,cm^{-3}$ where the quoted
uncertainty is the $1\sigma$ variation. 

\begin{figure}
\plotone{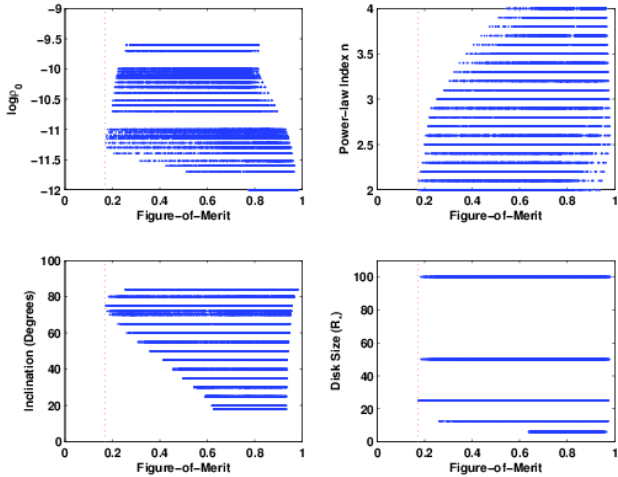}
\caption{\small The discrimination in the H$\alpha$ profile-fitting of
  the model parameters $\rho_0$, $n$, $i$ and $R_d$. Each horizontal
  line represents the figures-of-merit (Eq.~\ref{eq:fom}) for all models sharing that
  single parameter value.
\label{fig:discrim}}
\end{figure}

The influence of the model parameters $\rho_0$, $n$, $i$ and $R_d$ on
the line profile figure-of-merit ${\cal F}$ is shown in
Figure~\ref{fig:discrim} which plots each parameter versus ${\cal F}$
in a separate panel.  This results in a series of horizontal lines for
each parameter value (reflecting the discrete values of that parameter
considered) with the leftmost value giving the smallest ${\cal F}$
achievable with that choice. For example, for $i=20^{\circ}$, no
combination of the remaining model parameters $\rho_0$, $n$, and $R_d$
can result in ${\cal F}<0.6$. However, models near $i=75^{\circ}$
produce the overall minimum in ${\cal F}$ of around 0.18.  As
illustrated in the figure, the observed line profiles discriminate
most strongly in inclination, followed by $\rho_0$, and $n$. The
variation of ${\cal F}$ with $R_d$ rules out small disks with $R_d
\lesssim 5\,R_*$, but the discrimination for larger disk is poor as
these disks typically encompass the complete H$\alpha$ formation
region, and larger disks have little impact on the relative H$\alpha$
flux profile.

\begin{figure}
\plotone{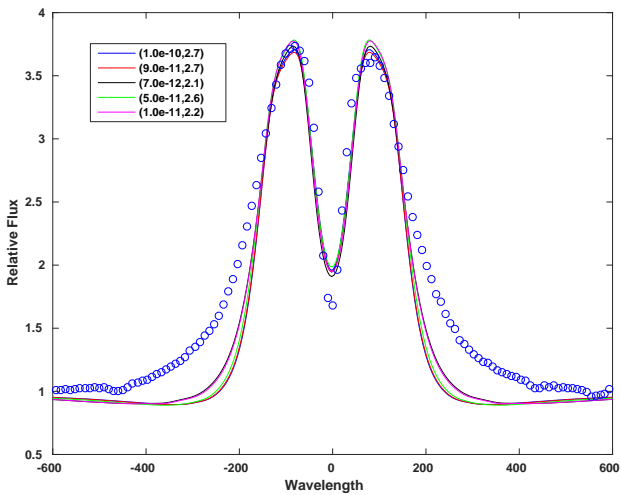}
\caption{\small The five best-fitting H$\alpha$ line profiles according
  to the core-weighted figure-of-merit given by Eq.~\ref{eq:fom2} (coloured lines)
  to the observed June 26, 2005 H$\alpha$ line profile (blue circles). 
  The disk parameters for each of the models (base disk density and power-law index)
  are as indicated in the legend. All models had $i=75^{\circ}$ and $R_d=25\,R_*$.
  \label{fig:bestfitHa_wgt}}
\end{figure}

The inability of our models to fit the H$\alpha$ line wings and core
simultaneously suggests that
we attempt to minimize the influence of the line wings on the fitting procedure in 
order to gauge the effect on the models selected. To this
end, we considered a revised, core-weighted, figure-of-merit, ${\cal F_{\rm cw}}$, of the form
\begin{equation}
{\cal F_{\rm cw}} \equiv \frac{1}{N}\sum_i w_i\,\frac{|F_i^{\rm mod} 
                - F_i^{\rm obs}|}{F_i^{\rm obs}} \;,
\label{eq:fom2}
\end{equation}
where the weights were chosen to be small in line wings and large in the core.
The function
\begin{equation}
w_i \equiv \frac{F_i^{\rm obs}}{F_c^{\rm obs}}-1 \;,
\end{equation}
where $F_c^{\rm obs}$ is the observed continuum flux (i.e.\ equal to unity as the spectra
are continuum normalized) achieves this effect.
Minimizing ${\cal F_{\rm cw}}$ results in best-fit profiles
that better fit the line core and the top five such profiles are shown in
Figure~\ref{fig:bestfitHa_wgt}. Interestingly, the top model parameters
now favour a higher $\rho_0$ and larger $n$ compared to the $w_i=1$ minimization.
The best fit to the June, 26, 2005 profile is $\rho_0=1.0\times 10^{-10}\,\rm g\,cm^{-3}$
and $n=2.7$ for $R_d=25\,R_*$ and $i=75^{\circ}$. Fitting all available spectra
from 2005 through 2014, and choosing the top-fitting profile in each case, gives essentially identical 
parameters except that $\rho_0$
varies from $5.0\times 10^{-11}$ through $1.0\times 10^{-10}\,\rm g\,cm^{-3}$. The average,
weighted by the number of fitting spectra, is $\rho_0=(6.8\pm 0.2)\,\times 10^{-11}\,\rm g\,cm^{-3}$,
nearly a factor of ten larger than the previous best-fit profiles based on the $w_i=1$
figure-of-merit.

Figure~\ref{fig:modelEW} also shows the selected models in the $(n,\log\rho_0)$ plane
that fit within 10\% of the best-fit model; the core-weighted fits include a much wider range of 
models.\footnote{In contrast to Eq.~(\ref{eq:fom}), including models within 20\% or
30\% of the best model does not significantly increase the range of models in 
the $(n,\log\rho_0)$ plane.} 
In conclusion, while this new weighting is arbitrary, it will become instructive when
we discuss the fit to the observed near-IR SED of o~Aqr in Section~\ref{sec:SED}.

\subsection{NPOI H$\alpha$ Interferometry}

\subsubsection{Geometric Models}
\label{sec:geo}

We first fit two very simple geometric models to the entire set of 994
NPOI visibilities: a nearly unresolved star (represented by a
uniform disk with an angular diameter of $0.222\;$mas) plus
either a uniform elliptical or Gaussian elliptical disk
representing the circumstellar contribution. If $V_*$ is the visibility of the
star and $V_D$ that of the disk, then the model visibilities can be
represented as
\begin{equation}
V^2=\left[ c_*\,V_*(0.222\,{\rm mas}) + (1-c_*)\,V_D(a,b,\phi)\right]^2 \;.
\label{eq:fitfun}
\end{equation}
Here, the major and minor axis of the ellipse ($a$ and $b$), the
position angle of the major axis ($\phi$), and the fractional
contribution of the star to the visibilities, $0<c_*<1$, are free
parameters. Detailed forms for $V_*$ and $V_D$ for both uniform and
Gaussian elliptical disks are given in \cite{ber03} and \cite{tyc06}.

Table~\ref{tab:geosummary} gives the results of these fits.  The
star-plus-elliptical Gaussian disk fits the observations with a
reduced chi-squared of $\chi^2/\nu=1.101$, yielding a major
axis~(given by the FWHM of the Gaussian) of $2.65\,\pm\,0.09\,$mas (or
equivalently a radial extent given by half-width at half-maximum of
$11.9\,R_*$). However, the fit is unconstrained along the minor axis,
consistent with $o$~Aqr being unresolved in this dimension, and no
estimate of the axial ratio is possible with this model.  The position
angle on the sky (measured East from North) is
$\phi=107\,\pm 6^{\circ}$.

The uniform elliptical disk model produces a slightly poorer fit
($\chi^2/\nu=1.139$) and a significantly larger major axis,
$4.15\,\pm\,0.15\,$mas, as expected based on the geometrically different
description of the extent of the emitting region~(diameter of uniform
disk versus FWHM of a Gaussian).  The axial ratio is found to be
$r=0.20\,\pm\,0.21$, again consistent with the minor axis not being
sufficiently resolved. The position angle on the sky for this model
was found to be $\phi=111\,\pm\,5^{\circ}$. Finally, we note that
assuming a geometrically thin disk, an axial ratio of $r=0.20\,\pm\,0.21$
implies a viewing inclination of $i=78\pm\,12^{\circ}$, consistent with
$i=75^{\circ}$ found from the H$\alpha$ line profile modelling of 
Section~\ref{subsec:halpha}.

\begin{figure}
\plotone{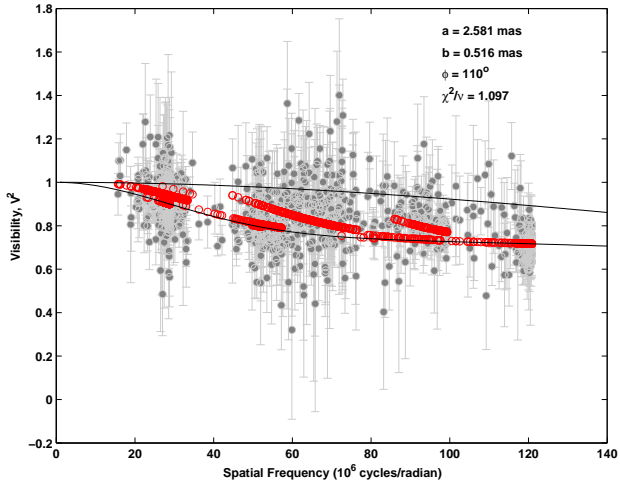}
\caption{\small The best-fitting star-plus-elliptical Gaussian disk
  geometric model. The axial ratio is fixed at $r=0.2$, as discussed
  in the text. The red symbols give the
  predicted model visibilities at the spatial frequencies of the
  observations, which are shown in light grey with $1\sigma$ error bars. 
  The visibilities of the major and minor axis of the
  model are shown as the solid lines.
\label{fig:bestellipse}}
\end{figure}

A star-plus-elliptical Gaussian disk model with a fixed axial ratio
of $r\equiv 0.2$ was also tried, and this gave a reduced chi-squared
of $\chi^2/\nu=1.097$, just marginally better than the unconstrained
fit. With this model, the major axis was $2.58\,\pm\,0.09\,$mas
and the position angle on the sky, $\phi=110\,\pm\,2^{\circ}$.
Figure~\ref{fig:bestellipse} compares the visibilities of this
star-plus-Gaussian disk with the fixed axial ratio of $r=0.2$ to the
NPOI observations.

In the fitting procedure, $c_*$ was treated as a free parameter with all
models finding $c_*=0.87$. However, as noted by \cite{tyc06}, 
$c_*$ is essentially 
\begin{equation}
c_*=\frac{\Delta}{\Delta+\rm EW_{H\alpha}} \,
\end{equation}
where $\Delta$ is the width of the NPOI filter, 150\,\AA, and $\rm
EW_{H\alpha}$ is the equivalent width of the H$\alpha$
emission. Using $\rm EW_{H\alpha} = 19.9\,$\AA, we find $c_*=0.88$, in
excellent agreement with the value recovered by the fits.

\citet{tou13} found a major axis of $1.525\pm 0.642\,$mas, 
an axial ratio of $r=0.249\pm 0.059$, and a
position angle of $\phi=107.5\pm2.2^{\circ}$ for $o$~Aqr based on K-band
continuum interferometry and Gaussian elliptical fits that included the
star. Their best-fit model had $\chi^2/\nu=1.80$. This
K-band continuum major axis found by \citet{tou13} is about 50\% smaller
than the H$\alpha$ major axis given in Table~\ref{tab:geosummary}.

$o$~Aqr
was also observed with VLT/AMBER by \citet{mei12} in both the K-band continuum and
Br$\gamma$. While o~Aqr was unresolved in the K-band, the Br$\gamma$
observations were consistent with a kinematic model with $i=70\pm 20^{\circ}$,
a position angle of $\phi=120\pm 20^{\circ}$, and a disk FWHM of $14\pm 1$ stellar
diameters. \citet{mei12} note sparse $(u,v)$ plane coverage and low S/N due to 
poor weather conditions. Nevertheless, the system inclination and position angle
of the major axis agree with geometric models of Table~\ref{tab:geosummary}.
One interesting comparison with the current work is that the Br$\alpha$ disk FWHM
found by \citet{mei12} is comparable to the H$\alpha$ disk FWHM found in this
work, something unusual for Be stars where the size of the H$\alpha$ region is
usually 1.5 to 2 times the Br$\gamma$ region. Similar sizes for these two
regions is further reflected by the very similar peak separations in the emission
profiles: $177\,\rm km\,s^{-1}$ for Br$\gamma$ \citep{mei12} and $162\,\rm km\,s^{-1}$
for H$\alpha$ (Figure~\ref{fig:halpha}).

Finally, we note that \citet{yud01} finds an intrinsic polarization
in the V~Band of $0.6$\% for o~Aqr with a position angle on the sky of 
$+6^{\circ}$. As the polarization vector is expected to be perpendicular 
to the major axis of the disk, this is consistent with the position angles 
found in Table~\ref{tab:geosummary}.

\begin{table*}[t]
\begin{center}
\small
\parbox{5.5in}{
\caption{Geometric model fits to the NPOI visibilities\label{tab:geosummary}}
} \\
\medskip
\begin{tabular}{@{}llllll}
\hline
\hline
Model         & Major Axis      & Axial Ratio       & Position Angle  & $c_*$  & $\chi^2/\nu$  \\
Star+         & (mas)           &                   & (deg)           &        &               \\
\hline
Uniform  Disk  & $4.15\,\pm 0.15$ & $0.20\,\pm 0.21$  & $111\,\pm\,5$  & $0.873\,\pm\,0.002$ & $1.139$ \\
Gaussian Disk  & $2.65\,\pm 0.10$ & $0^a$             & $107\,\pm\,6$  & $0.870\,\pm\,0.003$ & $1.101$ \\
Gaussian Disk  & $2.58\,\pm 0.09$ & $\equiv 0.2^b$    & $110\,\pm\,2$  & $0.864\,\pm\,0.003$ & $1.097$ \\
\hline
\end{tabular} \\
\medskip
$^a$ The minor axis is unconstrained by the fit. \\
$^b$ The axial ratio is fixed at the UD result.
\end{center}
\end{table*}

\subsubsection{Physical Models}
\label{sec:phys}

While the geometric fits of the previous section provided very good
representations of the observed visibilities, they cannot constrain
physical conditions in the disk, such as its density structure.  To
analyze the NPOI visibilities with physically-based models, the {\sc
  beray} code \citep{sig11} was used to produce images of the Be
star+disk models on the sky given the observer's viewing inclination.
Each model is specified by a choice of four parameters, ($\rho_0$,
$n$, $i$, and $R_d$), as with the H$\alpha$ spectroscopic
calculations.  The H$\alpha$ image computed by {\sc beray} was
integrated over a 150\,\AA\ wavelength interval centred on H$\alpha$
to match the NPOI observations.

Given the H$\alpha$ fit results of the previous section, {\sc beray}
images were computed for disk models with $\rho_o$ ranging from
$2.5\times 10^{-12}$ to $1.0\times 10^{-10}\,\rm g\,cm^{-3}$ and
power-law indexes $n=2.0$, $2.25$, $2.5$, $2.75$, and $3.0$.
Inclinations between $68$ to $80^{\circ}$ (in steps of $2^{\circ}$)
and $R_d=25$ and $50\,R_*$ were used. This subset of models includes
the entire region of the best-fit H$\alpha$ profiles.

Given a computed image specified by sky intensities $I_{ij}$, where
$i=1\ldots\,N_{x}^{\rm sky}$ and $j=1\ldots\,N_{y}^{\rm sky}$, the
predicted visibilities were determined by computing the discrete
Fourier transform of the image following the Zernike-van~Cittert
theorem \cite[e.g.\ see][]{lab06}.  In practice, the {\sc beray} image
was calculated with constant grid spacing on the sky within a linear
region spanning $R < 20\,R_*$ and a logarithmically-spaced grid beyond
that to reduce the computation time. To prepare for the DFT of the
image, the outer, non-linearly spaced region was interpolated down to
the constant spacing of the inner region.  As this interpolation is
done far away from the star and the image at these locations is
smooth, linear interpolation is sufficient. All images used a final
constant spacing of $0.05\;\rm R_*$ or $0.16\;\rm R_{\odot}$.
Before the DFT was computed, the image was zero-padded out to
$R=62.5\,R_*$ or $200\,R_{\odot}$.  The final images had $N_{x}^{\rm
sky}=N_{y}^{\rm sky}=2504$. The small grid spacing of the models
gives a Nyquist frequency of $\sim 1.8\times 10^{10}\,$ cycles per radian,
far larger than largest spatial frequency sampled by NPOI, and large
enough so that visibility values can be expected to be negligible even
for the nearly unresolved central star.

To compare with the observed visibilities, two-dimensional interpolation was performed
in the DFT images at each of the observed spatial frequencies $(u,v)$
of Figure~\ref{fig:uv}. To fit the position angle of the disk, it proved
more computationally efficient to rotate the $(u,v)$ coordinates of the
observed spatial frequencies, as opposed to the image itself.\footnote{We
thank Ludwik Lembryk for this suggestion.} The minimum in the reduced
$\chi^2$ defined the best position angle for a given image, and the
minimum in reduced $\chi^2$ over all trial images at their best-fit
position angle was used to define the best model.

\begin{figure}
\plotone{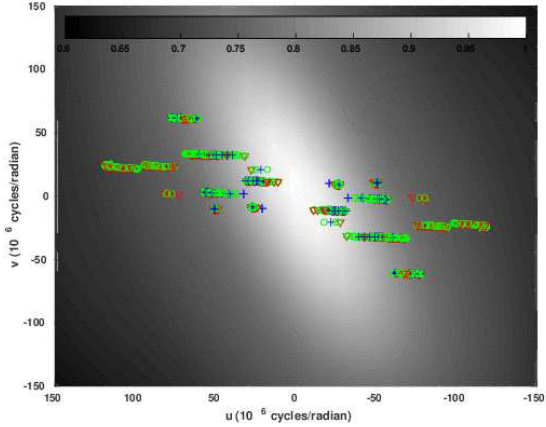}
\caption{\small The predicted visibilities based on the Fourier
  transform of the best-fit model image corresponding to the disk model 
  $\rho_0=5.0\times 10^{-12}\,\rm g\,cm^{-3}$, $n=2.0$, $R_d=25\,R_*$,
  and $i=78^{\circ}$ (grayscale, scale at top), compared to the observed
  visibilities (symbols). The position angle of the model is
  $110^{\circ}$. The symbols are: green circle (model fits
  to within error bars), red triangle (model below the
  observations), and blue plus sign (model above the
  observations). The overall reduced $\chi^{2}$ of the fit is 1.081
  for $N=994$.
\label{fig:bestDFT}}
\end{figure}

Over all trial images, the minimum reduced $\chi^2$ was found to be
$\chi^2/\nu=1.081$ ($N=994$), corresponding to the model with
$\rho_o=5.0\times 10^{-12}\,\rm g\,cm^{-3}$, $n=2.0$, $R_d=25\,R_*$.
Both the $i=78$ and $i=80^{\circ}$ images corresponding to this model
fit the data equally well, and the position angle of the major axis of
these best-fit models were $106^{\circ}$ and $110^{\circ}$. Thus the
physical {\sc bedisk}/{\sc beray} models are able to fit the observed
NPOI visibilities down to the limit set by the observational
uncertainties and slightly better than the geometric models of
Table~\ref{tab:geosummary}.  The comparison of the predicted
visibilities of this best-fit model to the observed NPOI visibilities
is shown in Figure~\ref{fig:bestDFT}.  Also, the best-fit visibility
model is in very good agreement with the previous best-fit model for
the H$\alpha$ line profile. Table~\ref{tab:physsummary} summarizes the
best-fit model parameters.

\begin{figure}
\plotone{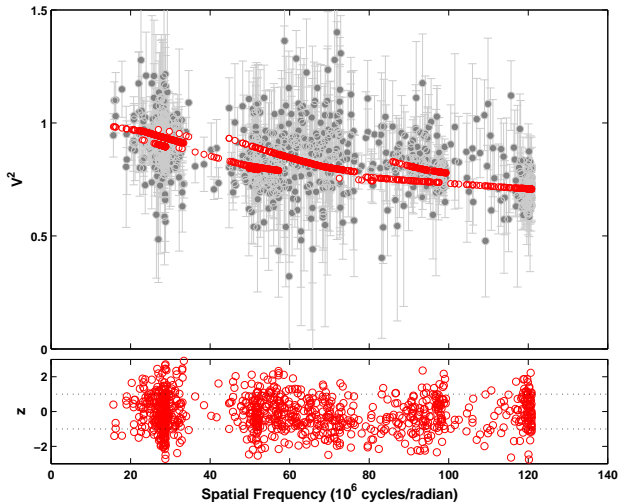}
\caption{\small The NPOI observations with errors compared with the
  model visibilities based on the Fourier transform of the best-fit model image
  with disk parameters
  $\rho_0 = 5.0\times 10^{-12}\,\rm g\,cm^{-3}$ and
  $n = 2.0$ as a function of the magnitude of the spatial frequency (top
  panel). The fit residuals, $z=(V^2_{\rm obs} - V^2_{\rm
    mod})/\sigma$, are shown in the lower panel.
\label{fig:datamodel}}
\end{figure}

\begin{figure}
\plotone{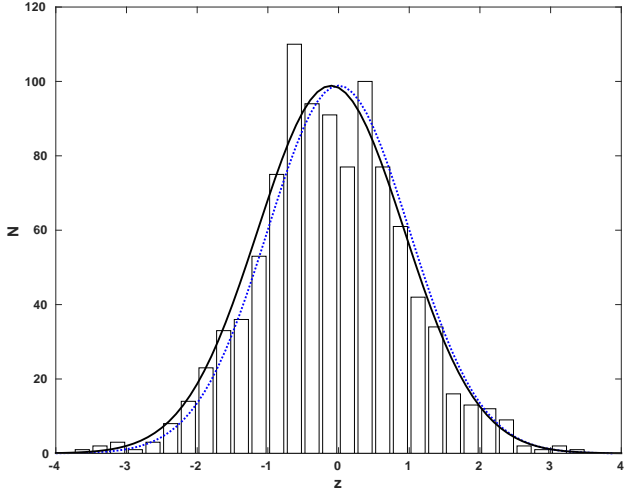}
\caption{\small Histogram of the visibility residuals, 
  $z=(V^2_{\rm model}-V^2_{\rm obs})/\sigma$, for the
  best-fit physical model. The solid black line shows a
  Gaussian fit to the residuals, giving $\mu=-0.1065$ and
  $\sigma=1.0372$. The dotted blue line gives
  the reference $\mu=0.0$, $\sigma=1$ Gaussian.
\label{fig:bestRes}}
\end{figure}

%
%
Figure~\ref{fig:datamodel} compares the observed visibilities
with the best-fit
disk model visibilities as a function of the magnitude of the spatial
frequency. Overall, 687 of the visibilities overlap the model values
within $\pm 1\sigma$. This results are consistent with that expected for 994
observations and 1$\sigma$ (or 68\%) error bounds.  The fit residuals,
defined as $z\equiv (V^2_{\rm obs} - V^2_{\rm mod})/\sigma$, are shown
as a function of spatial frequency in the lower panel of
Figure~\ref{fig:datamodel} and as a histogram in
Figure~\ref{fig:bestRes}. The best-fit Gaussian to the residual distribution
gives a mean of $\mu=-0.107\,\pm 0.065$, and a standard
deviation of $\sigma=1.037\pm 0.045$. A one-sample KS test on the
cumulative distribution of the residuals and that expected from the $N(-0.171,1.037)$
distribution gives a p-value of $0.686$, indicating the distribution is well-fit
by the Gaussian. While the standard deviation of the
residual distribution is consistent with 1, its mean is not consistent with 0.
This small shift of the residuals could possibly be eliminated by further refining
our models (i.e., finer grids in the model $\rho_0$ and $n$ disk density parameters),
but this would not lead to much additional insight.


\begin{figure}
\plotone{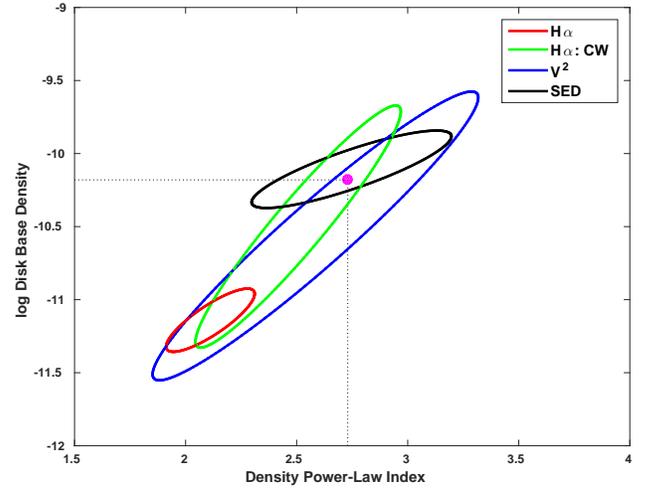}
\caption{\small The best-fit regions in the $(n,\log\rho_0)$ plane based on
the top 10\% of fits to the H$\alpha$ emission
profile~(red, Eq.~\ref{eq:fom}; green, Eq.~\ref{eq:fom2}), NPOI visibilities~(blue), and
\cite{tou10} SED~(black). The models enclosed for each feature correspond to the
`Notes' column in Table~\protect\ref{tab:physsummary}. The location of the adopted, best-fit
model for o~Aqr is shown as the purple circle.
\label{fig:summary10p}}
\end{figure}

Figure~\ref{fig:summary10p} shows the location in the $(n,\log\rho_o)$
plane of the best-fitting models based on constraints placed by
the H$\alpha$ emission profile, NPOI visibilities, and near-IR SED
(discussed in the next section).  The extent of the regions selected
in the $(n,\log\rho_o)$ plane reflects the fact that a number of models
produce similarly good fits to both the visibilities and the H$\alpha$
profile. For the visibilities, 104 models give a reduced $\chi^2$
within 10\% of the best fit, although it should be kept in mind that
the 560 models used in the visibility analysis were deliberately chosen
as a plausible sub-sample of models based on the previous fits to the
H$\alpha$ emission line profile, and many of these models differ only
in their viewing inclination angle over the range $68$ to $80^{\circ}$.
The wide range of models consistent with the visibilities includes the
smaller ranges consistent with the H$\alpha$ profile at the same level.


Note that the diagonal trend to the best-fitting models in the
($n,\log\rho_0$) diagram is expected and has been noted before
\citep{tyc08}. For an optically thick disk, the flux is, to first
order, just the Planck function at the average disk temperature times
the projected surface area of the disk out to $\tau=1$. Therefore, the
various combinations of $\rho_0$ and $n$ that produce similar
effective emitting areas will result to first order in observational
signatures that match the observations equally well.

\begin{table*}[t]
\begin{center}
\small
\parbox{5.5in}{
\caption{Summary of Best-Fit Physical Models\label{tab:physsummary}}
} \\
\medskip
\begin{tabular}{@{}lllllll}
\hline
\hline
Feature Used to & Best Fitting  & $\rho_0$ & $n$ & $R_d$  & $i$ & Notes \\
Constrain Fit   & Parameter     & ($\rm g\,cm^{-3}$) &     & ($R_*$) & (degrees) &       \\
\hline
H$\alpha^a$& ${\cal F}=1.85\times 10^{-1}$   & $5.0\times 10^{-12}$ & $2.0$ & 25 & 75$^o$ & 17 models within 10\% \\
H$\alpha^a$ core-weighted& ${\cal F}_{\rm CW}=3.5\times 10^{-4}$   & $1.0\times 10^{-10}$ & $2.7$ & 25 & 75$^o$ & 10 models within 10\% \\
V$^2$ Visibility   & $\chi^2/\nu=1.081$  & $5.0\times 10^{-12}$ & $2.0$ & 25 & $78-80^o$ & 104 models within 10\% \\
Near-IR SED        & $\chi^2/\nu=0.296$  & $1.0\times 10^{-10}$ & $3.0$ & 25 & 72$^o$ & 17 models with $\chi^2/\nu<1$ \\
                   &                     &                      &       &    &        &  \\
Adopted      & Region in Figure~\ref{fig:summary10p} & $6.6\times 10^{-11}$ & $2.7$ & 25 & 75$^o$ &  \\
\hline
\end{tabular} \\
\end{center}
\medskip
$^a$ The reported parameters are for the June, 2005 profile. \\
\end{table*}


\begin{figure}
\plotone{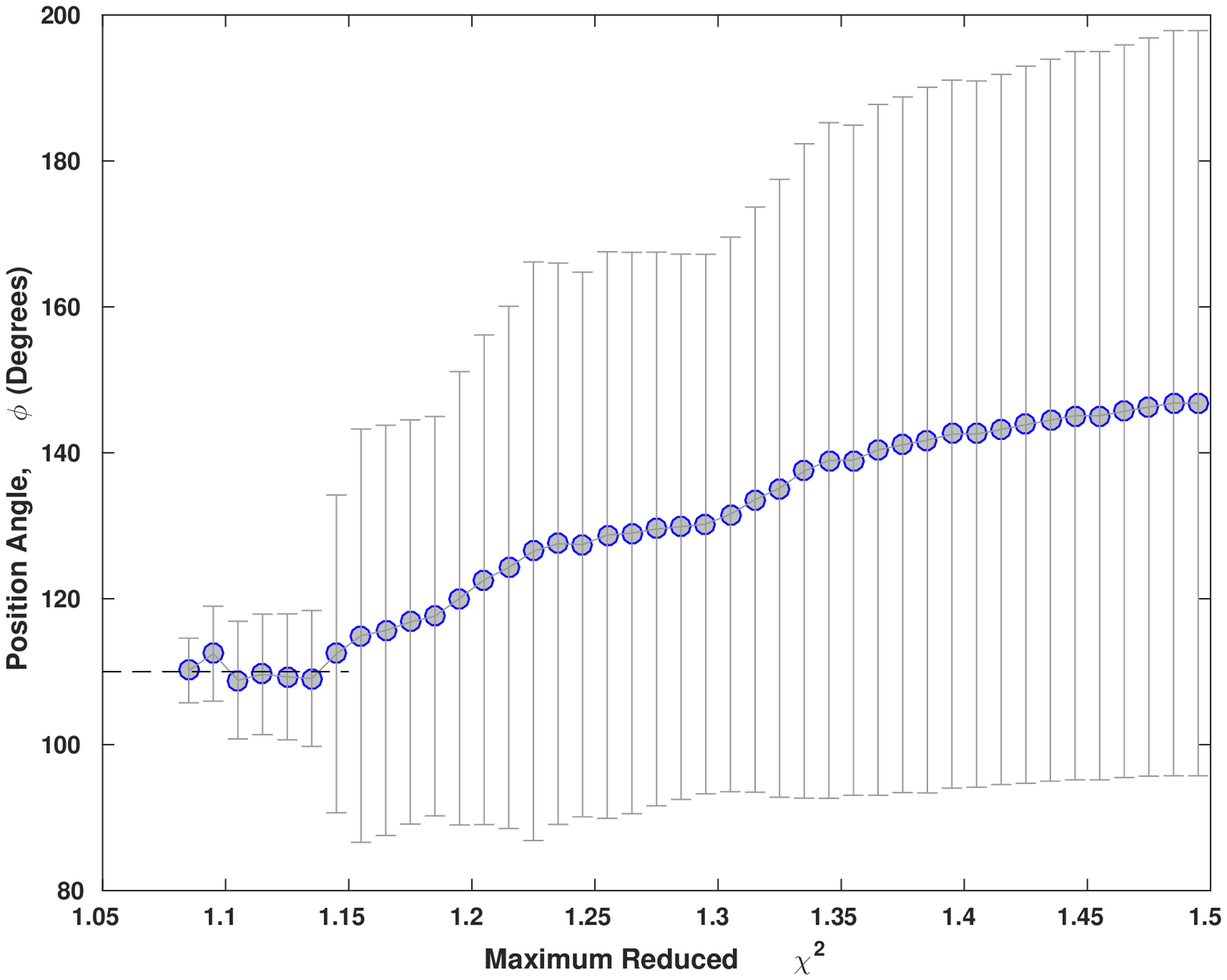}
\caption{\small The mean position angle $\phi$ and its $1\sigma$ variation for all
models with a reduced $\chi^2$ less than a given maximum. A position angle of
$\phi=110^o$ is shown by the horizontal dotted line, and it fits the mean position
angle over all models with a reduced $\chi^2 < 1.15$.
\label{fig:pos_v_fom}}
\end{figure}

Figure~\ref{fig:pos_v_fom} shows the individual best-fit position angles averaged over all trial
images with a reduced $\chi^2$ less than a given threshold. The error bars
are the $1\sigma$ variation about this mean. For the best-fitting images,
$\chi^2/\nu < 1.15$, the means are all consistent with
$\phi=110 \pm 8^o$ which we adopt as the best estimate of the position angle
of o~Aqr's disk on the sky as derived from the NPOI visibility data. 
This value is consistent with the position angles found from the elliptical
uniform disk and Gaussian geometric models (see Table~\ref{tab:geosummary}) which 
fit the visibility data almost as well as the physical model discussed here.

Figure~\ref{fig:pos_v_fom} also shows that as worse-fit
models are included in the average, the mean position angle rises steadily,
and the $1\sigma$ variation increases dramatically. Finally, we note that for
individual images, the discrimination in position angle is good as the
reduced $\chi^2$ varies by more than factor of two over the range in
$\phi$ from 0 to $180^{\circ}$.

Given the time-span of the NPOI observations, we have also fit subsets
of the visibility data.  As expected, the more limited 2007 data are
consistent with a significantly larger range of physical parameters. We
did not analyze the 2011 data separately as it is confined to smaller
spatial frequencies (see Figure~\ref{fig:v2}), while analyzing the
2012-2014 data alone gives results that are indistinguishable from the
full data set.  Given this, it is not possible to detect variability
between the 2007 and 2012-2014 data sets.


%
%

\subsection{The Near-IR SED}
\label{sec:SED}

\citet{tou10} give optical and near-infrared fluxes (from 2008) for
$o$~Aqr at four wavelengths: $\lambda\,0.440$, $0.680$, $1.654$, and
$2.179\,\rm \mu m$.  The apparent stability of $o$~Aqr's disk suggests
that it is useful to consider these near-IR fluxes as an additional
consistency check on our modelling. {\sc beray} was used to compute
optical and near-IR spectral energy distributions for the same subset
of disk models used to analyze the NPOI visibilities.

To compare to the model fluxes, the \citet{tou10} observations were
normalized to the model SED at $0.440\,\rm\mu m$, and the reduced $\chi^2$
of the fit computed for the three remaining observed wavelengths. The
uncertainties in the observed fluxes given by \citet{tou10} are typically
$<10\%$, and are quoted as the quadratic sum of uncertainties due
to instrumental error, errors due to repeatability of the individual
observations, and errors associated with the calibration and air-mass
corrections

\begin{figure}
\plotone{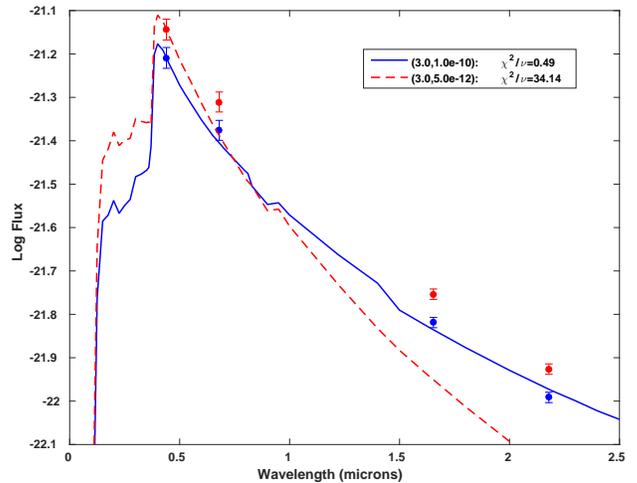}
\caption{\small Comparison of the model SEDs with near-IR measurements
of \protect\cite{tou10}. The best-fit model SED, with $n=3.0$,
$\rho_0=10^{-10}\,\rm g\,cm^{-3}$, $R_d=25\,R_*$ and $i=72^{\circ}$,
is shown as the solid blue line.
The worst-fit SED model with the same $R_d$ and $i$ is shown as the
dotted red line. The reduced-$\chi^2$ of each model is given in the caption.
Note that the observations have been separately normalized to each model's prediction 
at $0.44\,\mu$m.  \label{fig:touhami}}
\end{figure}

Figure~\ref{fig:touhami} shows the best fit to the observed near-IR SED.
This model has $\rho_o =1.0\times 10^{-10}\,\rm g\,cm^{-3}$, $n=3.0$, $R_d=25\,R_*$,
and $i=72^{\circ}$, and fits the observations well with a $\chi^2/\nu=0.49$.
Also shown in the figure is the worst-fitting model that has the
same $R_d$ and $i$. This is a lower-density model, 
$\rho_o =5.0\times 10^{-12}\,\rm g\,cm^{-3}$ and $n=3.0$, which gives fluxes close to the 
pure photospheric SED of the star alone. As the observed fluxes have been
separately normalized to each model prediction at $0.440\,\rm\mu m$, the model
fluxes themselves can be directly compared to each other. This illustrates that the best-fit model predicts
an IR-excess and optical and UV deficits relative to the photospheric
spectrum. The small deficits of $\leq\sim 0.1\,$mag are a consequence of the obscuration
of the photosphere by the disk for the large viewing inclination.

In addition to this best-fit model, 17 of the
560 models considered had a reduced $\chi^2$ of less than unity. All models
consistent with the \citet{tou10} SED are represented by the black ellipse
in the $(n,\log\rho_0)$ plane shown in Figure~\ref{fig:summary10p}.

Interestingly, the near-IR SED is consistent only with the
H$\alpha$ line profile fitting when the core-weighted figure of
merit is used, Equation~\ref{eq:fom2}. This situation is summarized
in Figure~\ref{fig:summary10p}. All three contemporaneous
observational constraints, the H$\alpha$ line profile (fit using
core-weighting), the NPOI visibilities, and the \citet{tou10} near-IR
SED, imply a best-fit model of $\rho_o =6.6\times 10^{-11}\,\rm
g\,cm^{-3}$, $n=2.7$ with $R_d=25\,R_*$ seen at an inclination of
$i=75^{\circ}$. These will be adopted as the disk parameters for o~Aqr
over the time period considered.



Figure~\ref{fig:summary10p} also shows that the disk parameters
of the best-fit, uniformly weighted H$\alpha$ profile (i.e., fit with
Eq.~\ref{eq:fom}) are not consistent with those based on the near-IR
SED of \citet{tou10}. Nevertheless, the spatial extent of the H$\alpha$ disk is
much larger than that contributing to the near-IR flux. The extended wings
of H$\alpha$ are not fully reproduced by any of our model H$\alpha$ profiles,
and this may reflect additional material close to the star not accounted for in
our assumption of a single power-law description of radial density fall-off.

\subsection{Disk Density Variations:}

%
%

\begin{figure}
\plotone{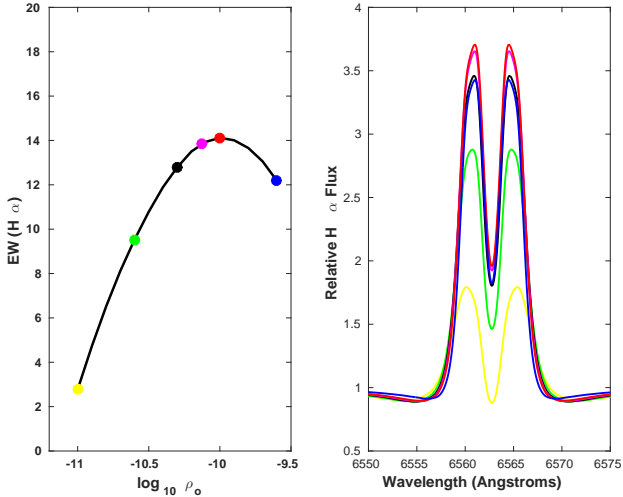}
\caption{\small Left panel: the H$\alpha$ equivalent width as a function of 
$\rho_0$ for the model $n=2.7$, $R_d=25\,\rm R_*$ seen at an inclination of 
$i=75^{\circ}$. Right panel: the H$\alpha$ line profile as a function
of increasing $\rho_0$ for the same model. The $\rho_0$ for each of the five depicted profiles 
corresponds to the colour of the filled circle in the left panel. \label{fig:eqwrho}}
\end{figure}

The consistency of the H$\alpha$ equivalent width and line profile seen in
Figure~\ref{fig:halpha} suggests that the density in $o$~Aqr's disk is
very stable over this time period. However it is important to understand the possible
limitations of using H$\alpha$ as a proxy for disk density stability.
This is illustrated in Figure~\ref{fig:eqwrho} which shows the equivalent width
(left panel) and line profile (right panel) of H$\alpha$ as a function of the disk
base density $\rho_0$ for the model with $n=2.7$, $R_d=25$ and $i=75^o$.
The maximum predicted equivalent width is 14.1\,\AA\ at $\log\rho_0=-10.0$.  Near
this maximum, both the H$\alpha$ profile and equivalent width are very insensitive 
to changes in the disk density: for example, as the density increases from
$5.0\times\,10^{-11}\,\rm g\,cm^{-3}$, the maximum increase in the equivalent width
is about 10\% before returning to the starting value by $2.5\times\,10^{-10}\,\rm g\,cm^{-3}$.
The line profiles for $5.0\times\,10^{-11}\,\rm g\,cm^{-3}$ and $2.5\times\,10^{-10}\,\rm g\,cm^{-3}$
(black and blue profiles in Figure~\ref{fig:eqwrho}, respectively)
are virtually identical.  Thus, in this range of disk parameters, small variations in 
either the equivalent width or profile of H$\alpha$ can mask large changes in the disk density.

In the current case of o~Aqr, it is significant that disk a density model consistent with
all considered observational constraints (H$\alpha$ profile fit with core-weighting, the visibilities, and
the near-IR SED) with parameters $\rho_0=6.7\times\,10^{-11}\,\rm g\,cm^{-3}$ and $n=2.7$ is very near the 
maximum predicted model strength. This provides a natural explanation for the observed stability of the H$\alpha$
line profile and equivalent width. As noted above, even large changes in the overall
disk density, up to a factor of approximately five, will lead to only small changes in
the observed H$\alpha$ profile.


\subsection{The CQE Feature in Mg\,{\sc ii} $\lambda\,4481$}
\label{sec:cqe}

\cite{riv06} note that $o$~Aqr exhibits a central quasi-emission (CQE)
feature in the core of Mg\,{\sc ii} $\lambda\,4481$. CQE features are
apparent emission ``bumps" in the cores of some lines, particularly
those of Mg\,{\sc ii}, He\,{\sc i}, and Fe\,{\sc ii}. Despite their
appearance as relative emission, CQEs are a pure absorption effect
caused by the velocity shift (in the observer's frame) of the local
atomic line profile in a Keplerian-rotating disk viewed nearly edge-on
\citep{han96}. Because of this geometrical requirement, and the somewhat
special circumstances of their formation, CQEs can be a useful test of
a particular disk model. In this section, we show that the appearance
of a CQE feature in Mg\,{\sc ii} $\lambda\,4481$ is consistent with the
disk density parameters found for $o$~Aqr in this work.

\begin{figure}
\plotone{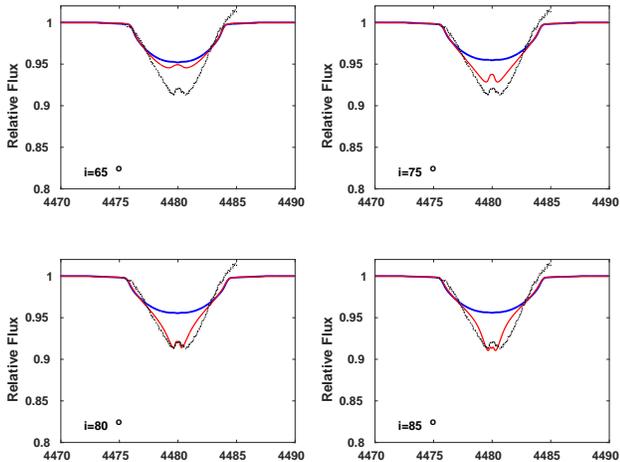}
\caption{\small Mg\,{\sc ii} $\lambda\,4481$ line profiles for disk
  models with $\rho_0=1\times 10^{-11}\,\rm g cm^{-3}$ (blue) or
  $\rho_0=5\times 10^{-11}\,\rm g cm^{-3}$ (red) and $n=2.5$,
  $R_d=25\,R_*$ and the viewing inclination indicated in each
  panel. The 1999 observations of \protect\cite{riv06} are shown
  as the black dotted profile.
\label{fig:mgII}}
\end{figure}

\cite{riv06} present a Mg\,{\sc ii} $\lambda\,4481$ profile from 1999
(somewhat outside of the time-frame of the present work) which shows a CQE
feature with a central amplitude of just less than 1\% ($F_c/F_m=1.008$
where $F_c$ is the line centre flux and $F_m$ is the flux minimum
just outside the core; see Figure~\ref{fig:mgII}). To see if this
is consistent with the disk model proposed for $o$~Aqr, we have used
{\sc beray} to compute Mg\,{\sc ii} $\lambda\,4481$ line profiles
for circumstellar disks with $\rho_0$ values of $5\times 10^{-12}$,
$10^{-11}$, $5\times 10^{-11}$ and $10^{-10}\;\rm g\,cm^{-3}$, all
with $n=2.5$, $R_d=25\,R_*$, for viewing inclinations of $i=65^{\circ}$,
$75^{\circ}$, $80^{\circ}$ and $85^{\circ}$. We have assumed an equatorial
velocity of $290\;\rm km s^{-1}$ for $o$~Aqr, corresponding to $v_{\rm
frac}=0.74$ \citep{tou13}. Photospheric profiles were computed assuming
LTE as this is a reasonable approximation at the $T_{\rm eff}$ of
$o$~Aqr \citep{sig95}.

Figure~\ref{fig:mgII} shows that for $\rho_0=5\times 10^{-11}\,\rm g\,
cm^{-3}$, a CQE feature of the correct amplitude is predicted for
$i=75^{\circ}$ and $80^{\circ}$. Interestingly, 
a CQE feature is not predicted for densities less than $\rho_0=10^{-11}\;\rm g cm^{-3}$. 
Thus the appearance of a CQE feature in Mg\,{\sc ii} $\lambda\,4481$ is
consistent with the proposed disk density model found in this work
and shown in Figure~\ref{fig:summary10p}. Note that as these observations
are outside time-frame considered in this work, we have not attempted to find
a best-fit profile to Mg\,{\sc ii} $\lambda\,4481$. Nevertheless, the general agreement
we find is a non-trivial test of our proposed disk density model for o~Aqr.


%

%
%

\subsection{The Mass and Angular-Momentum Content of the H$\alpha$ Disk}
\label{sec:ml}

Combining H$\alpha$ spectroscopy, NPOI interferometric
visibilities, and the near-IR SED, we determined a best-fit disk model of $\rho_o=6.6\times
10^{-11}\,\rm g\,cm^{-3}$, $n=2.7$, $R_d=25\,R_*$ seen at an
inclination of $i=75^{\circ}$. The position angle of the major axis on
the sky was found to be $110\,\pm\,8^{\circ}$.

To determine the mass in the H$\alpha$ disk implied by this model,
we have computed an additional image for the best-fit parameters but for
$i=0^{\circ}$. Plotting $2\pi R\,I$ versus $R$, where $I$ is the model
intensity at distance $R$, we find that $90\%$ of the H$\alpha$ disk light
comes from $R\leq 9.5\,R_*$. Using the best-fit disk parameters for
$(n,\rho_0)$ above and the disk density model given by Eq.~(\ref{eq:rho}),
we find an enclosed disk mass of $1.5\times 10^{24}\,\rm g$ or $\sim
1.8\times 10^{-10}\,M_*$.

Assuming Keplerian rotation for the disk, well established for Be
stars \citep{riv13a}, we find a total angular momentum associated
with the H$\alpha$ emitting disk of $3.5\times 10^{43}\,\rm
g\,cm^{2}\,s^{-1}$. The stellar angular momentum is $J_*=\beta^2\,M_* R_*
V_{eq}$ where $\beta$ is the radius of gyration, $\sim 0.2$ \citep{cla89},
and $V_{eq}$ is $o$~Aqr's equatorial velocity, $290\;\rm km\,s^{-1}$. We
find $J_*=2.2\times 10^{51}\,\rm g\,cm^2\, s^{-1}$, making the angular
momentum associated with the H$\alpha$ emitting disk $\sim 1.6\times
10^{-8}\,J_{*}$.

To give an indication of the robustness of these values for the total
mass and angular momentum content of the H$\alpha$ emitting disk, we list
in Table~\ref{tab:mass_summary} the mass and angular momentum values for the best-fit
disk models for each of the considered observational constraints separately, as given
in Table~\ref{tab:physsummary}. The disk radius
that encloses 90\% of the H$\alpha$ emission is computed for each model and
is given in the Table~\ref{tab:mass_summary}.  The range is about a factor of three both in disk mass and five in 
angular momentum content. This is significantly smaller than the range in the disk base density $\rho_0$ 
alone as higher densities are associated with larger values for the power-law index $n$.

\begin{table*}[t]
\begin{center}
\small
\parbox{5.5in}{
\caption{Summary of disk mass and total angular momentum content.\label{tab:mass_summary}}
} \\
\medskip
\begin{tabular}{@{}lrllll}
\hline
\hline
Diagnostic & $(R_{90}/R_*)^a$  & $M_d$     & $M_d/M_*$    & $J_d$    & $J_d/J_*$ \\
           &                   & (gm)      &              & ($\rm g\,cm^2\,s^{-1}$)  &   \\
\hline
H$\alpha$ (${\cal F}$)          & 19.5 & $1.0\times 10^{+24}$  & $1.2\times 10^{-10}$  & $3.7\times 10^{+43}$  & $1.7\times 10^{-8}$ \\
H$\alpha$ (${\cal F}_{\rm CW}$) &  8.3 & $2.0\times 10^{+24}$  & $2.3\times 10^{-10}$  & $4.4\times 10^{+43}$  & $2.0\times 10^{-8}$ \\
$V^2$                           & 19.5 & $1.0\times 10^{+24}$  & $1.2\times 10^{-10}$  & $3.7\times 10^{+43}$  & $1.7\times 10^{-8}$ \\
near-IR SED                     &  3.3 & $5.8\times 10^{+23}$  & $6.9\times 10^{-11}$  & $9.0\times 10^{+42}$  & $4.2\times 10^{-9}$ \\
                                &              &              &              & \\
Adopted$^b$                     &  9.5 & $1.5\times 10^{+24}$  & $1.8\times 10^{-10}$  & $3.5\times 10^{+43}$  & $1.6\times 10^{-8}$ \\ \hline
\end{tabular} \\
\end{center}
\medskip
$^a$ $R_{90}$ is the disk radius that encloses 90\% of the integrated H$\alpha$ light.\\
\medskip
$^b$ This is the best-fit model to all three constraints (H$\alpha$ (${\cal F}_{\rm CW}$),
$V^2$, and near-IR SED) shown in Figure~\ref{fig:summary10p}.
\end{table*}

%
%
\section{Conclusions}
\label{sec:discuss}

We have analyzed a large set ($N=994$) of H$\alpha$ interferometric
visibilities obtained from the Navy Precision Optical Interferometer
(NPOI) for the Be shell star $o$~Aqr, covering the period 2007 through
2014. Using predicted visibilities based on physical disk models
computed by the {\sc bedisk} and {\sc beray} codes, we find best-fit
disk parameters that are consistent with an analysis of the H$\alpha$
emission line profile and the near-IR SED of \citet{tou10} from
the same time period. We note that these physically-based {\sc beray} 
images can fit the observed visibilities down to the level associated with 
observational uncertainties~(i.e., down to the level of $\chi^2/\nu\sim 1$).

The best-fit disk model with $\rho_o=6.6\times 10^{-11}\,\rm g\,cm^{-3}$,
$n=2.7$, $R_d=25\,R_*$, implies a disk mass associated with the
H$\alpha$ emitting region of $\sim 1.8\times 10^{-10}\,M_*$ and an
angular momentum content of the disk of $\sim 1.6\times 10^{-8}\,J_{*}$,
where $M_*$ and $J_*$ are the mass and angular momentum of the central
$B$ star in $o$~Aqr.

Over the nine years of H$\alpha$ spectroscopic observations, from 2005
until 2014, we  find variations in its equivalent width of typically
less than 5\%. However our best-fit, disk density model is at the maximum
strength of H$\alpha$ that can be produced for any value of $\rho_0$ given
a power-law index of $n=2.7$ seen at $i=75^{\circ}$. For this model,
variations in the disk density, $\rho_0$, of up to a factor of $\sim 5$
would not lead to noticeable changes in the H$\alpha$ equivalent width or
its profile and variations of this magnitude cannot be excluded
over the time period considered.

We further test our model by comparing to the 1999 CQE feature in
Mg\,{\sc ii} $\lambda\,4481$ observed by \cite{riv06} and find that
very similar feature can be produced by our best-fit disk density
model, representing an additional  and highly non-trivial success of our
modelling.

Finally, we note that in order to produce fits to the H$\alpha$
line profile consistent with the other constraints, the fits needed to
be weighted more heavily in the line core (emission peaks and central
depression) at the expense of the line wings.
This may reflect the limitation of the
assumption of a single power-law for the disk's equatorial density.

%
%

\vspace{0.3in} 
\small

We would like to thank the anonymous referee and Doug Gies for many
helpful comments.  This work is supported by the Canadian Natural Sciences and
Engineering Research Council (NSERC) through a Discovery Grant to TAAS
and by CMU through a FRCE Type~B grant to CT.  We would like to thank
the NPOI project staff for their support in acquiring the
interferometric data used in this work, and Brennan Kerkstra and
Sandeep Chiluka for assistance with NPOI data reductions.  The Navy
Precision Optical Interferometer is a joint project of the Naval
Research Laboratory and the US Naval Observatory, in cooperation with
Lowell Observatory and is funded by the Office of Naval Research and
the Oceanographer of the Navy.  We thank the Lowell Observatory for
the telescope time used to obtain the H$\alpha$ line spectra presented
in this work.

{\it Facilities:} \facility{NPOI}, \facility{Hall (Solar Stellar Spectrograph)}

\end{document}